\documentclass{article}
\usepackage{spie}
\usepackage{epsfig}
\usepackage{graphicx}
\usepackage{subfigure}
\usepackage{floatflt}
\title{Quantum harmonic oscillator state\\
 synthesis and analysis\footnote{~~Work of the U.S.\ Government.
 Not subject to U.S.\ copyright.}}

\author{W. M. Itano, C. Monroe, D. M. Meekhof,\\ D. Leibfried,
B. E. King, and D. J. Wineland
\skiplinehalf
Time and Frequency Division\\
National Institute of Standards and Technology\\
Boulder, Colorado \hspace{0.5em}80303 \hspace{0.5em}USA
}

\authorinfo{Send correspondence to W.M.I.
E-mail: witano@nist.gov; telephone: 303-497-5632; fax: 303-497-6461
}

\begin{document}

\maketitle


\begin{abstract}
We laser-cool single beryllium ions in a Paul trap to the ground
$(n=0)$ quantum harmonic oscillator state with greater than 90\%
probability.
From this starting point, we can put the atom into various
quantum states of motion by application of optical and rf electric
fields.
Some of these states resemble classical states (the coherent
states), while others are intrinsically quantum, such as number
states or squeezed states.
We have created entangled position and spin superposition states
(Schr\"odinger cat states), where the atom's spatial wavefunction
is split into two widely separated wave packets.
We have developed methods to reconstruct the density matrices
and Wigner functions of arbitrary motional quantum states.
These methods should make it possible to study decoherence of
quantum superposition states and the transition from quantum
to classical behavior.
Calculations of the decoherence of superpositions of coherent
states are presented.
\end{abstract}



\section{INTRODUCTION} 

In a series of studies
we have prepared single, trapped, $^9$Be$^+$ ions in various quantum
harmonic oscillator states and performed measurements on those states.
In this article, we summarize some of the results.
Further details are given in the original
reports \cite{monroe95,meekhof96,monroe96,leibfried96,leibfried97}.

The quantum states of the simple harmonic oscillator have been
studied since the earliest days of quantum mechanics.
For example, the harmonic oscillator was among the first applications
of the matrix mechanics of Heisenberg \cite{heisenberg25}
and the wave mechanics of Schr\"odinger \cite{schrodinger26}.
The theoretical interest in harmonic oscillators is partly due to
the fact that harmonic oscillator problems
often have exact solutions.
In addition, physical systems, such as vibrating molecules,
mechanical resonators, or modes
of the electromagnetic field, can  be modeled as harmonic oscillators,
so that the theoretical results can be compared to experiments.

A single ion in a Paul trap can be described effectively as
a simple harmonic oscillator, even though the Hamiltonian is actually
time-dependent, so  no stationary states exist.
For practical purposes, the system can be treated as if
the Hamiltonian were that of an ordinary, time-independent harmonic
oscillator (see, e.g., Refs.~\citenum{bardroff96,schrade95}).

\section{SYSTEM AND EFFECTIVE HAMILTONIAN}

The effective Hamiltonian for the center-of-mass secular
motion is that of an anisotropic
three-dimensional harmonic oscillator.
If we choose an interaction Hamiltonian which affects only the
$x$-motion, then we can deal with the one-dimensional harmonic
oscillator Hamiltonian,
\begin{equation}
H_{x}=\hbar\omega_x a_x^\dagger a_x,
\end{equation}
where $\omega_x$ is the secular frequency for the $x$-motion,
and $a_x^\dagger$ and $a_x$ are the creation and annihilation operators
for the quanta of the $x$-oscillation mode.
The constant term $\hbar\omega_x/2$, which results from the usual
quantization procedure, has been left out for convenience.
The eigenstates of $H_{x}$ are $|n_x\rangle$,
where
\begin{equation}
H_{x}|n_x\rangle=
n_x\hbar\omega_x|n_x\rangle.
\end{equation}

The internal states of the $^9$Be$^+$ ion which are the most important
for the experiments are shown in Fig.~\ref{fig:levels}.
We are mostly concerned with
the hyperfine-Zeeman sublevels of the ground $2s\,^2S_{1/2}$ electronic
state, which
are denoted by $|2s\,^2S_{1/2};F,M_F\rangle$, where ${\bf F}$ is the
total angular momentum, and $M_F$ is the eigenvalue of $F_z$.
The $^9$Be nucleus has spin 3/2.
Of chief importance are $|2s\,^2S_{1/2};2,2\rangle$, abbreviated as
$|\!\downarrow\rangle$, and $|2s\,^2S_{1/2};1,1\rangle$, abbreviated as
$|\!\uparrow\rangle$.
The hyperfine-Zeeman sublevels of the $2p\,^2P_J$
($J=1/2$ or $3/2$) fine-structure multiplet
also play a role, either as intermediate
states in stimulated Raman transitions or as the final states in
resonantly-driven single-photon transitions used for laser cooling
or state detection.
We denote these states by $|2p\,^2P_J;F,M_F\rangle$.
The energy separation of $|\!\uparrow\rangle$ and $|\!\downarrow\rangle$
is $\hbar\omega_0$, where $\omega_0\approx 2\pi\times 1.250$ GHz.
They are coupled by laser beams R1 and R2,
through the intermediate $|2p\,^2P_{1/2};2,2\rangle$ state.
The frequency detuning from the intermediate state is $\Delta$,
where $\Delta\approx -2\pi\times 12$ GHz.

If we restrict the internal states to the space spanned by
$|\!\uparrow\rangle$ and $|\!\downarrow\rangle$, then the internal
Hamiltonian can be written as
\begin{equation}
H_{\uparrow\downarrow}=\frac{\hbar\omega_0}{2}\sigma_z,
\end{equation}
where $\sigma_z$ is a Pauli spin matrix whose nonzero matrix elements
are$\langle \uparrow\!|\sigma_z|\!\uparrow\rangle=+1$
and $\langle \downarrow\!|\sigma_z|\!\downarrow\rangle=-1$.
%
\begin{floatingfigure}[l]{7cm}
\epsfxsize=6cm
\epsfbox{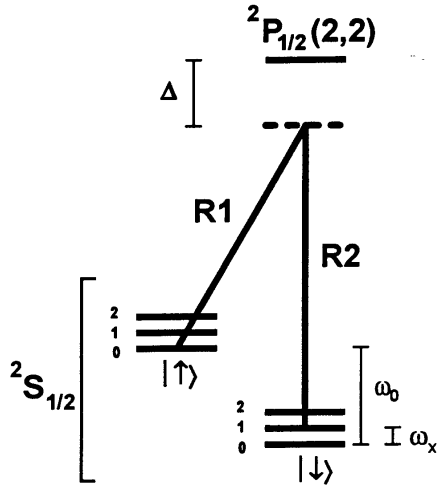}
\caption[levels]
{\label{fig:levels}
Internal and motional energy levels of a $^9$Be$^+$ ion.
The $|\!\uparrow\rangle$ and $|\!\downarrow\rangle$ states are particular
hyperfine-Zeeman components
of the ground $^2S_{1/2}$ state, separated in energy by
$\hbar\omega_0$.
Each internal state can exist in any of a ladder of vibrational
energy states $|n_x\rangle$, where $n_x=0,1,2,\ldots$,
separated in energy by $\hbar\omega_x$,
where $\omega_x\approx 2\pi\times 11.2$ MHz.
}
\end{floatingfigure}


The total effective Hamiltonian for the system consisting of the
internal $|\!\uparrow\rangle$ and $|\!\downarrow\rangle$ states
and the $x$ motional degree of freedom is
\begin{equation}
H=H_{\uparrow\downarrow}+H_x+H_{\rm int},
\end{equation}
where $H_{\rm int}$ is the effective interaction Hamiltonian coupling
due to the two laser beams R1 and R2 in Fig.~\ref{fig:levels}.
We define an interaction-picture operator ${\cal O}^I$ in terms
of a Schr\"odinger-picture operator ${\cal O}^S$ as
\begin{equation}
{\cal O}^I(t)=e^{iH_{\uparrow\downarrow}t/\hbar}{\cal O}^S(t)
e^{-iH_{\uparrow\downarrow}t/\hbar}.
\end{equation}
The effective interaction Hamiltonian in the interaction picture
and the rotating wave approximation is
\begin{equation}
\label{eq:effhamiltonian}
H^I_{\rm int}=\hbar g\left(\sigma_{+}e^{i\eta(a_x^\dagger+a_x)-i\delta t}
+\sigma_{-}e^{-i\eta(a_x^\dagger+a_x)+i\delta t} \right),
\end{equation}
where $g$ is the interaction strength, $\delta$ is the detuning of the
frequency difference of the two laser beams with respect to $\omega_0$,
and $\eta=k\sqrt{\hbar/(2m\omega_x)}$ is the Lamb-Dicke parameter,
where $k$ is the magnitude of the difference between the wavevectors of
the two laser beams, and $m$ is the mass of the ion.
The nonzero matrix elements of $\sigma_{+}$ and $\sigma_{-}$
are
$\langle \uparrow\!|\sigma_{+}|\!\downarrow\rangle=
\langle \downarrow\!|\sigma_{-}|\!\uparrow\rangle=1$.

The detuning $\delta$ can be tuned to multiples of $\omega_x$,
$\delta=(n^\prime -n)\omega_x$, so as to resonantly drive
transitions between $|\!\downarrow,n\rangle$ and
$|\!\uparrow,n^\prime\rangle$.
We refer to the $\delta=0$ resonance as the carrier, the
$\delta=-\omega_x$ resonance as the first red sideband,
and the $\delta=+\omega_x$ resonance as the first blue sideband.

The signal that is detected in the experiments is the
probability\newline
\hspace*{7.3cm} $P_\downarrow(t)$ that the ion is in the
$|\!\!\downarrow\rangle$
internal state after a particular preparation.
If we irradiate the ion at time $t$ with circularly polarized
light, resonant with the electronic transition
from $|\!\downarrow\rangle$ to $|2p\,^2P_{3/2};3,3\rangle$, there
will be a high fluorescence intensity if the $|\!\downarrow\rangle$
state is occupied, since the selection rules only allow that upper state
to decay back to the $|\!\downarrow\rangle$ state, so it can continue
to scatter photons.
This transition is called a cycling transition.
If the ion is in the $|\!\uparrow\rangle$ state when it is irradiated
with the same light, it will scatter a negligible number of photons.
Thus, if we repeatedly
prepare the ion in the same way, apply radiation
resonant with the cycling transition, and
detect the fluorescence photons, the average signal will be proportional
to $P_\downarrow(t)$.


\section{CREATION AND PARTIAL MEASUREMENTS OF QUANTUM STATES}
\label{sec:creation}

\subsection{Fock states}

The ion is prepared in the $n=0$ state by Raman cooling \cite{monroe95}.
(From now on we drop the $x$ label on $n$.)
Raman cooling consists of a sequence of laser pulses on the red sideband,
driving $|\!\downarrow,n\rangle$ to $|\!\uparrow,n-1\rangle$ transitions,
followed by laser pulses which recycle the ion back to the
$|\!\downarrow,n-1\rangle$ state.
The probability of heating due to recoil during the recycling step is small.
At the end of the sequence, the ion is in the $n=0$ state more than
90\% of the time.

A Fock state is another name for an $n$-state.
Higher-$n$ Fock states are prepared from the $n=0$ state by a sequence
of $\pi$-pulses on the blue sideband, red sideband, or carrier.
For example, the $|\!\uparrow,2\rangle$ state is prepared by using
blue sideband, red sideband, and carrier $\pi$-pulses in succession,
so that the ion steps through the states $|\!\downarrow,0\rangle$,
$|\!\uparrow,1\rangle$, $|\!\downarrow,2\rangle$,
and $|\!\uparrow,2\rangle$.

\begin{figure}[tbhp]
\begin{center}
\subfigure{
\includegraphics[width=8cm,clip]{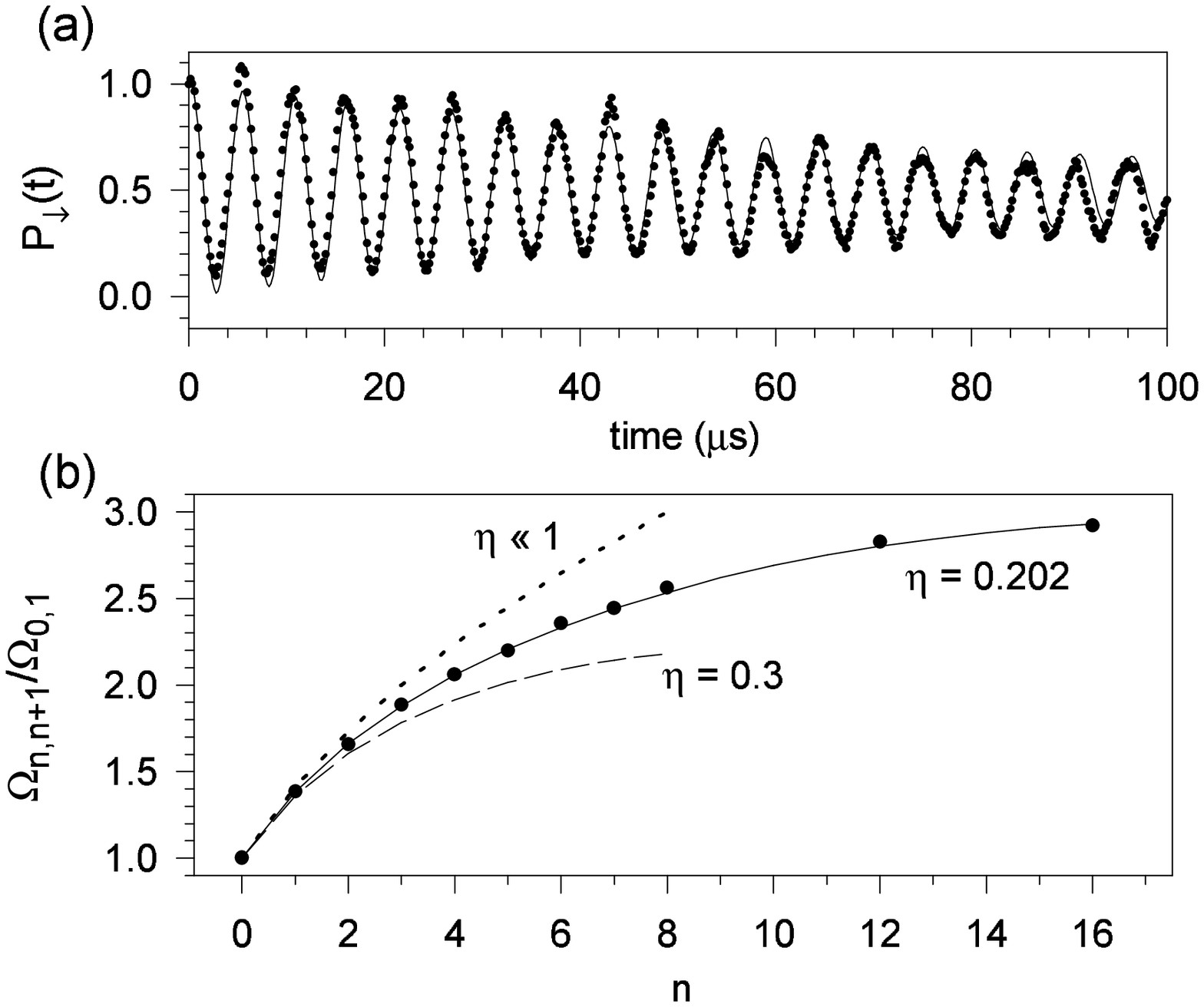}
}
\hspace{.5cm}
\subfigure{
\includegraphics[width=8cm,clip]{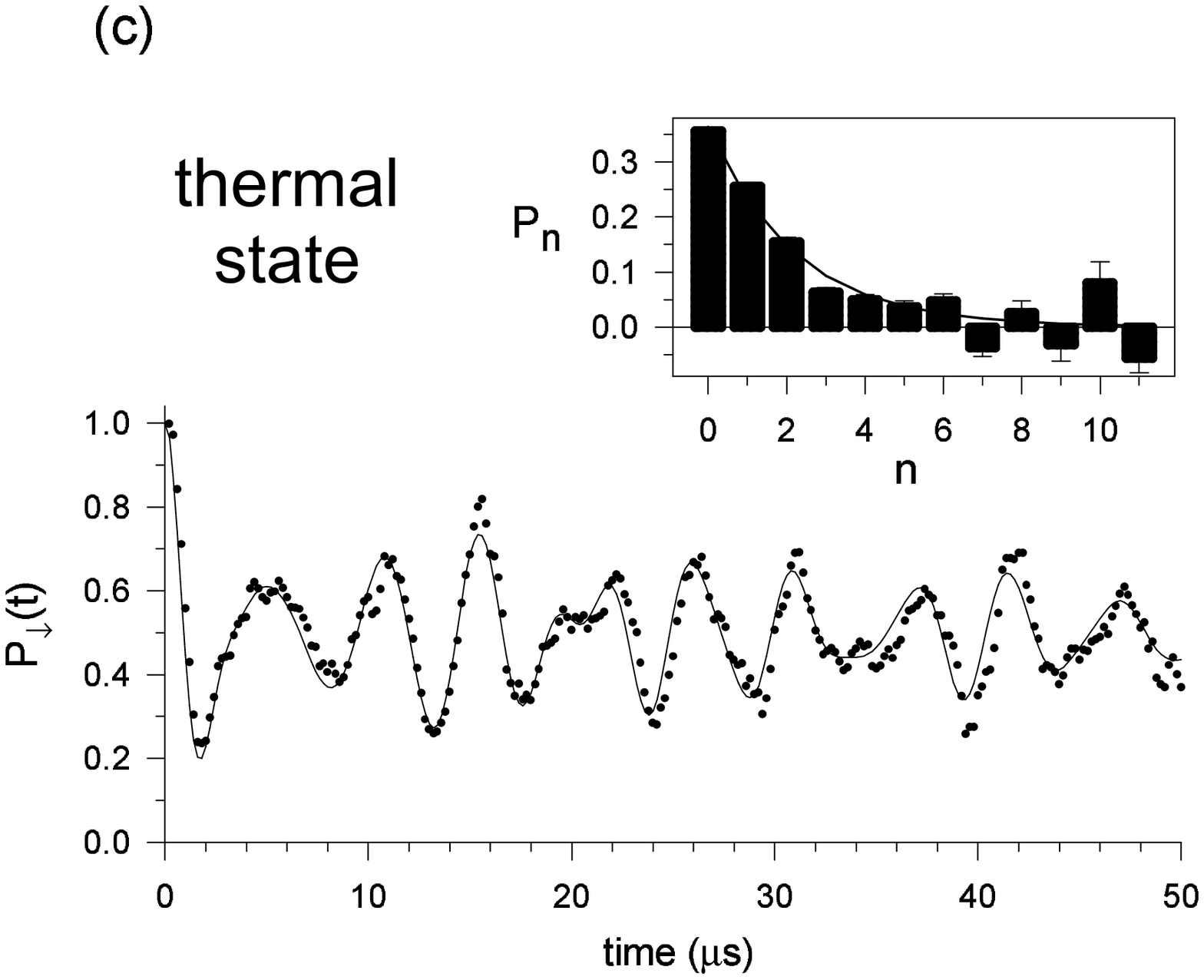}
}
\end{center}
\caption[focktrace]
{\label{fig:focktrace}
(a) $P_\downarrow(t)$ for an initial $|\!\downarrow,0\rangle$
number state driven by the first blue sideband.
The solid line is a fit to an exponentially damped sinusoid.
(b) Observed ratios of the Rabi frequencies $\Omega_{n,n+1}/\Omega_{0,1}$
for different values of the initial $n$.
The lines represent calculated values for different values of the
Lamb-Dicke parameter $\eta$.
The value of $\eta$, inferred from other measured
quantities, was $0.202\pm 0.005$.
(c)
$P_\downarrow(t)$ for a thermal state.
The solid line is a fit of the data (dots)
to a sum of Fock states having a thermal
distribution. The fitted value for the mean value of $n$ is
$\overline{n}=1.3\pm 0.1$.
The inset shows the amplitudes of the Fock state components (bars)
with a fit to an exponential, corresponding to $\overline{n}=
1.5\pm 0.1$ (line).
}
\end{figure}

If the atom is initially in the $|\!\downarrow,n\rangle$ state,
and the first blue sideband is driven,
it will oscillate between that state and the $|\!\uparrow,n+1\rangle$ state.
The probability of finding it in the $|\!\downarrow,n\rangle$ state at a time
$t$ is
\begin{equation}
\label{eq:rabiflopping}
P_\downarrow(t)=\frac{1}{2}\left[1+\cos(2\Omega_{n,n+1}t)
e^{-\gamma_n t}\right],
\end{equation}
where $\Omega_{n,n+1}$, the Rabi flopping rate, is a function of the
laser intensities and detunings and $n$, and $\gamma_n$ is a damping
factor which is determined empirically.
Thus, the frequency of the oscillations of $P_\downarrow(t)$ is
a signature of the initial value of $n$.
Figure~\ref{fig:focktrace}(a) shows $P_\downarrow(t)$ for an initial
$|\!\downarrow,0\rangle$ state.
Figure~\ref{fig:focktrace}(b) shows the observed ratios of Rabi
frequencies compared with the values calculated from the matrix elements
of the interaction Hamiltonian [Eq.~(\ref{eq:effhamiltonian})].


\subsection{Thermal states}
A thermal state of the motion of the ion is not a pure state,
but rather must be described by a density matrix, even though
there is only one ion.
The statistical ensemble which is described by the density matrix
is generated by repeatedly preparing the state and making a measurement.

The ion is prepared in a thermal state by Doppler laser cooling
\cite{stenholm86}.
The temperature of the distribution can be controlled by changing
the detuning of the cooling laser.
When the ion's state is not a Fock state, $P_\downarrow(t)$ has the
form
\begin{equation}
P_\downarrow(t)=\frac{1}{2}\left[ 1+\sum_{n=0}^\infty
P_n \cos(2\Omega_{n,n+1}t) e^{-\gamma_n t}\right],
\end{equation}
where $P_n$ is the probability of finding the ion in the state
$|n\rangle$.
Figure \ref{fig:focktrace}(c) shows $P_\downarrow(t)$ for a thermal
state.



\subsection{Coherent states}
\label{subsec:coherent}

\begin{figure}[tbhp]
\begin{center}
\subfigure[]{
\includegraphics[width=8cm,clip]{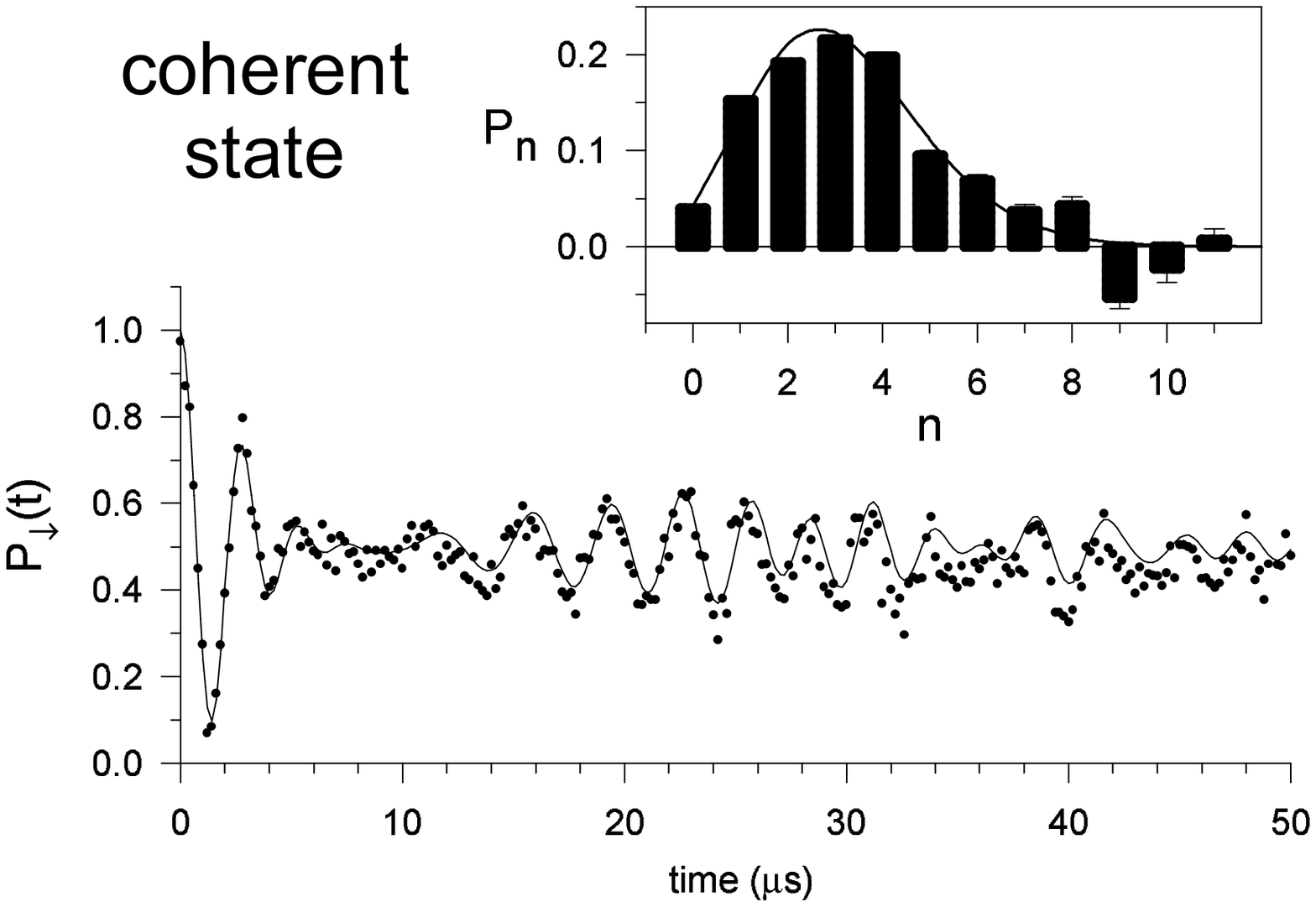}
}
\hspace{.5cm}
\subfigure[]{
\includegraphics[width=8cm,clip]{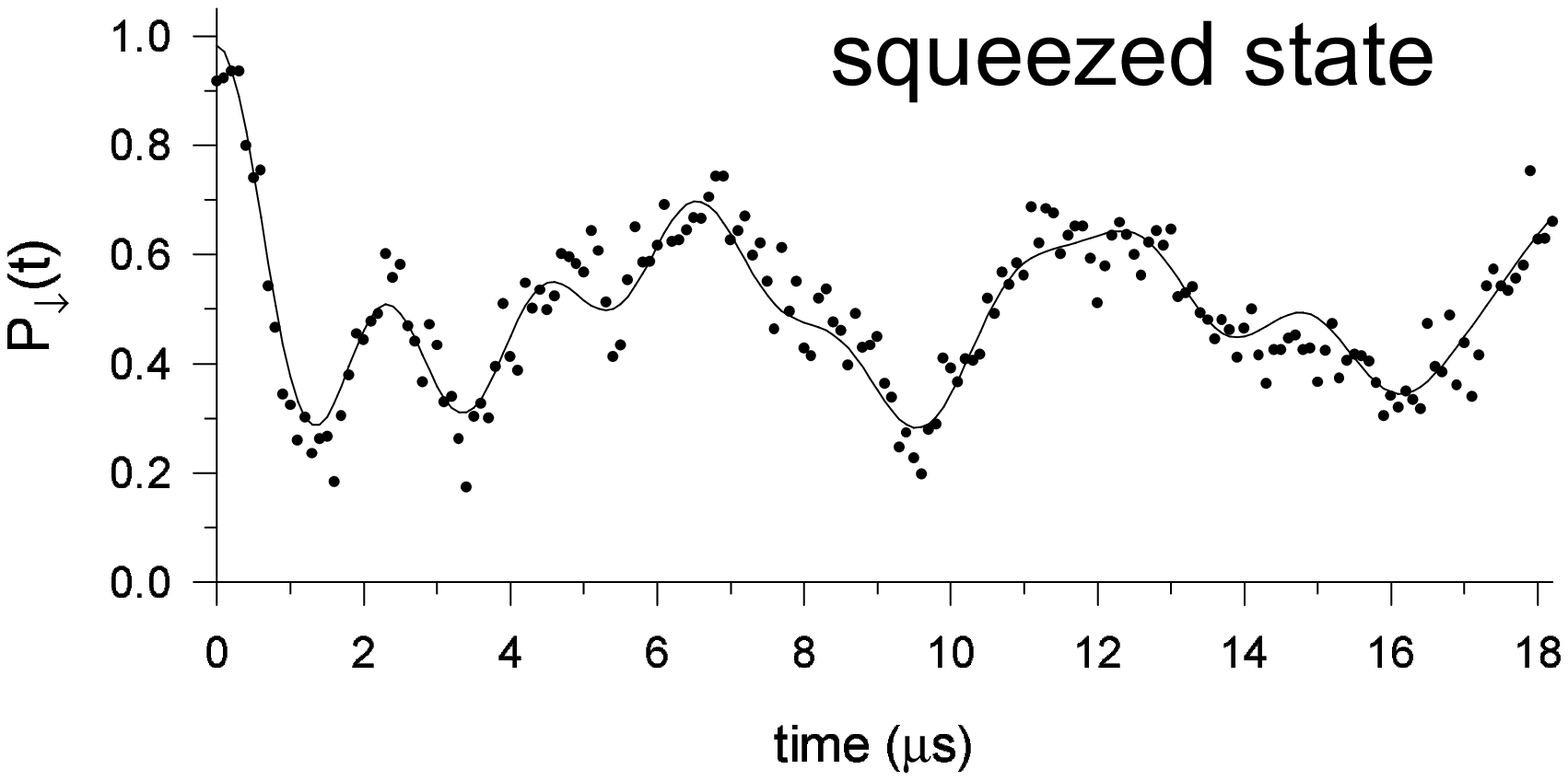}
}
\end{center}
\caption[coherenttrace]
{\label{fig:coherenttrace}
(a)
$P_\downarrow(t)$ for a coherent state.
The solid line is a fit of the data (dots)
to a sum of Fock states having a coherent state distribution.
The fitted value for $\overline{n}$ is $3.1\pm 0.1$.
The inset shows the amplitudes of the Fock state components (bars)
with a fit to a Poissonian distribution, corresponding to $\overline{n}=
2.9\pm 0.1$ (line).
(b)
$P_\downarrow(t)$ for a squeezed state.
The solid line is a fit of the data (dots)
to a sum of Fock states having a squeezed-state distribution.
The fitted value for $\beta$ is $40\pm 10$, which
corresponds to $\overline{n}\approx 7.1$
}
\end{figure}

Coherent states of the quantum harmonic oscillator were introduced
by Schr\"odinger \cite{schrodinger26a}, with the aim of describing
a classical particle with a wavefunction.
A coherent state $|\alpha\rangle$ is equal to the following superposition
of number states:
\begin{equation}
\label{eq:coherent-def}
|\alpha\rangle=e^{-\frac{1}{2}{|\alpha|}^2}\sum_{n=0}^\infty
\frac{\alpha^n}{\sqrt{n!}}|n\rangle.
\end{equation}
In the Schr\"odinger picture,
the absolute square of the wavefunction retains its shape, and
its center follows the trajectory of a classical particle in
a harmonic well.
The mean value of $n$ is $\overline{n}=|\alpha|^2$.
A coherent state can be created from the $n=0$ state (a special case of
coherent state) by applying a spatially uniform classical force
(see Appendix A).
The drive is most effective when its frequency is $\omega_x$.
Another method is to apply a ``moving standing wave,'' that is,
two  laser beams differing in frequency by $\omega_x$
and differing in propagation direction,
so that an oscillating dipole force is generated \cite{wineland92}.
We have used both of these methods to prepare coherent states.
Figure~\ref{fig:coherenttrace}(a) shows $P_\downarrow(t)$ for
a coherent state.
This trace exhibits the phenomenon of collapse and revival
\cite{eberly80}.

\subsection{Squeezed states}


\begin{figure}[tbhp]
\begin{center}
\includegraphics[width=10cm,clip]{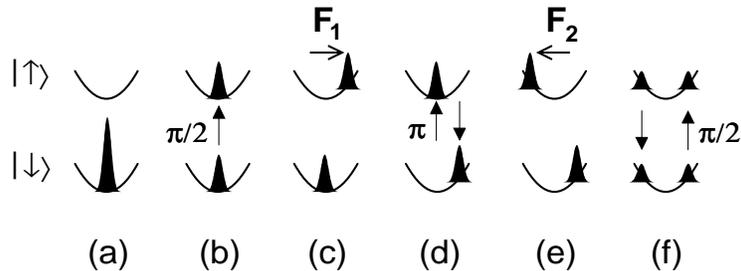}
\end{center}
\caption[catpic]
{\label{fig:catpic}
Creation of a Schr\"odinger cat state.
See text for details.
}
\end{figure}

A ``vacuum squeezed state'' can be created by applying an electric
field gradient having a frequency of $2\omega_x$ to an ion initially
in the $n=0$ state \cite{heinzen90}.
Here, the ion was irradiated with
two laser beams which differed in frequency by $2\omega_x$.
This has the same effect.
The squeeze parameter $\beta$ is defined as the factor by which the
variance of the squeezed quadrature is decreased.
It increases exponentially with the time the driving force is applied.
The probability distribution of $n$-states is nonzero only for even $n$,
for which,
\begin{equation}
P_{2n}=N\frac{(2n)!(\tanh r)^{2n}}{(2^n n!)^2},
\end{equation}
where $\beta=e^{2r}$, and $N$ is a normalization constant.
Figure~\ref{fig:coherenttrace}(b) shows $P_\downarrow(t)$ for a squeezed
state having $\beta=40\pm 10$.



\subsection{Schr\"odinger cat states}


\begin{figure}[htbp]
\begin{center}
\includegraphics[width=6cm,clip]{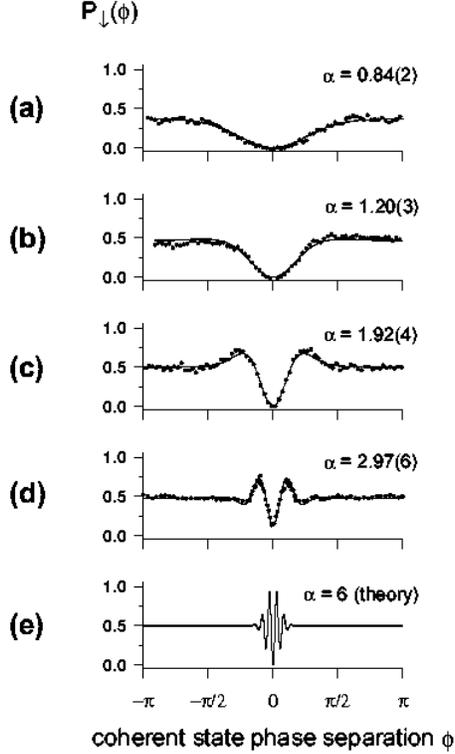}
\end{center}
\caption[catgraph]
{\label{fig:catgraph}
Measured and fitted $P_\downarrow(\phi)$ interference signals
for the Schr\"odinger cat state.
The curves in (a) through (d) represent measurements for
different values of the time
during which the displacement beams excite the coherent states.
The solid lines are fits to Eq.~(\protect\ref{eq:catsignal}),
allowing $\alpha$ to vary.
In (d), a term representing loss of contrast was included in the fit.
The curve in (d) represents a superposition of two $x_0\approx 7$~nm
wave packets with a maximum separation of $4\alpha x_0\approx 80$~nm.
Curve (e) represents a calculation for a pair of coherent states with
$\alpha=6$.
}
\end{figure}

The term ``Schr\"odinger cat'' is taken here to denote an entangled
state which consists of two coherent states of motion
correlated with different internal atomic states.
A simple example is
\begin{equation}
|\Psi\rangle=\frac{|\!\uparrow,\alpha\rangle+
|\!\downarrow,-\alpha\rangle}{\sqrt{2}}.
\end{equation}

Figure \ref{fig:catpic} shows how such a state is created.
(a) The ion is prepared in the $|\!\downarrow,0\rangle$ state
by Raman cooling.
(b) A $\pi/2$-pulse on the carrier generates an equal superposition
of $|\!\downarrow,0\rangle$ and $|\!\uparrow,0\rangle$.
(c) The displacement beams generate a force ${\bf F_1}$,
which excites the component in the
$|\!\uparrow\rangle$ internal state to a coherent state $|\alpha\rangle$.
Due to the polarizations of the displacement beams, they
do not affect an atom in the $|\!\downarrow\rangle$ state.
(d) A $\pi$-pulse on the carrier exchanges the $|\!\downarrow\rangle$
and $|\!\uparrow\rangle$ components.
(e) The displacement beams generate a force ${\bf F_2}$,
which excites the component in the
$|\!\uparrow\rangle$ internal state to a coherent state
$|\alpha e^{i\phi}\rangle$, where the phase $\phi$ is controlled
by an rf oscillator.
The state here is analogous to Schr\"odinger's cat.
(f) The $|\!\downarrow\rangle$ and $|\!\uparrow\rangle$ components
are combined by a $\pi/2$-pulse on the carrier.
At this point, the radiation is applied on the cycling transition,
and the signal is recorded.

The predicted signal, for a particular value of $\phi$, is
\begin{equation}
\label{eq:catsignal}
P_\downarrow(\phi)=\frac{1}{2}\left[1-
ce^{-\alpha^2 (1-\cos\phi)}\cos(\alpha^2\sin\phi)\right],
\end{equation}
where $c=1$ in the absence of decoherence.
Figure \ref{fig:catgraph} shows experimental data and fits to
Eq.~(\ref{eq:catsignal}) for various values of $\alpha$.
In (d), $c$ is clearly less than one, indicating decoherence,
although the source is not yet determined.


\section{COMPLETE MEASUREMENTS OF QUANTUM STATES}

The measurements described in the previous Section determine only the
$n$-state populations or probabilities, and therefore do not
provide a complete description of the motional states.
The density matrix $\rho$ does provide a complete description of
a state, whether it is a pure or a mixed state.
The Wigner function $W(\alpha)$ also provides a complete description.
The Wigner function resembles a classical joint probability distribution
for position and momentum in some cases.
However, it can be negative, unlike a true probability distribution.
We have demonstrated experimental methods for reconstructing the
density matrix or the Wigner function of a quantum state of motion
of a harmonically bound atom \cite{leibfried96}.

Both of these methods depend on controllably displacing the state
in phase space, applying radiation to drive the first blue sideband
for time $t$, and then measuring $P_\downarrow$.
The averaged, normalized, signal is
\begin{equation}
\label{eq:displaced-flop}
P_\downarrow(t,\alpha)=\frac{1}{2}\left[1+\sum_{k=0}^\infty
Q_k(\alpha)\cos(2\Omega_{k,k+1}t) e^{-\gamma_k t}
\right],
\end{equation}
where the complex number $\alpha$ represents the amplitude and phase
of the displacement,
and $Q(\alpha)$ is the occupation probability of the vibrational state
$|k\rangle$ for the displaced state.

If the $Q_k(\alpha)$ coefficients are derived for a series of values
of $\alpha$ lying in a circle,
\begin{equation}
\alpha_p=|\alpha|\exp[i(\pi/N)p],
\end{equation}
where $p=-N,\ldots,N-1$, then the density matrix elements $\rho_{nm}$
can be determined for values of $n$ and $m$ up to $N-1$.
The details of the numerical procedure are given in Ref.~\citenum{leibfried96}.

\begin{figure}[tbhp]
\begin{center}
\subfigure[]{\includegraphics[width=5cm,clip]{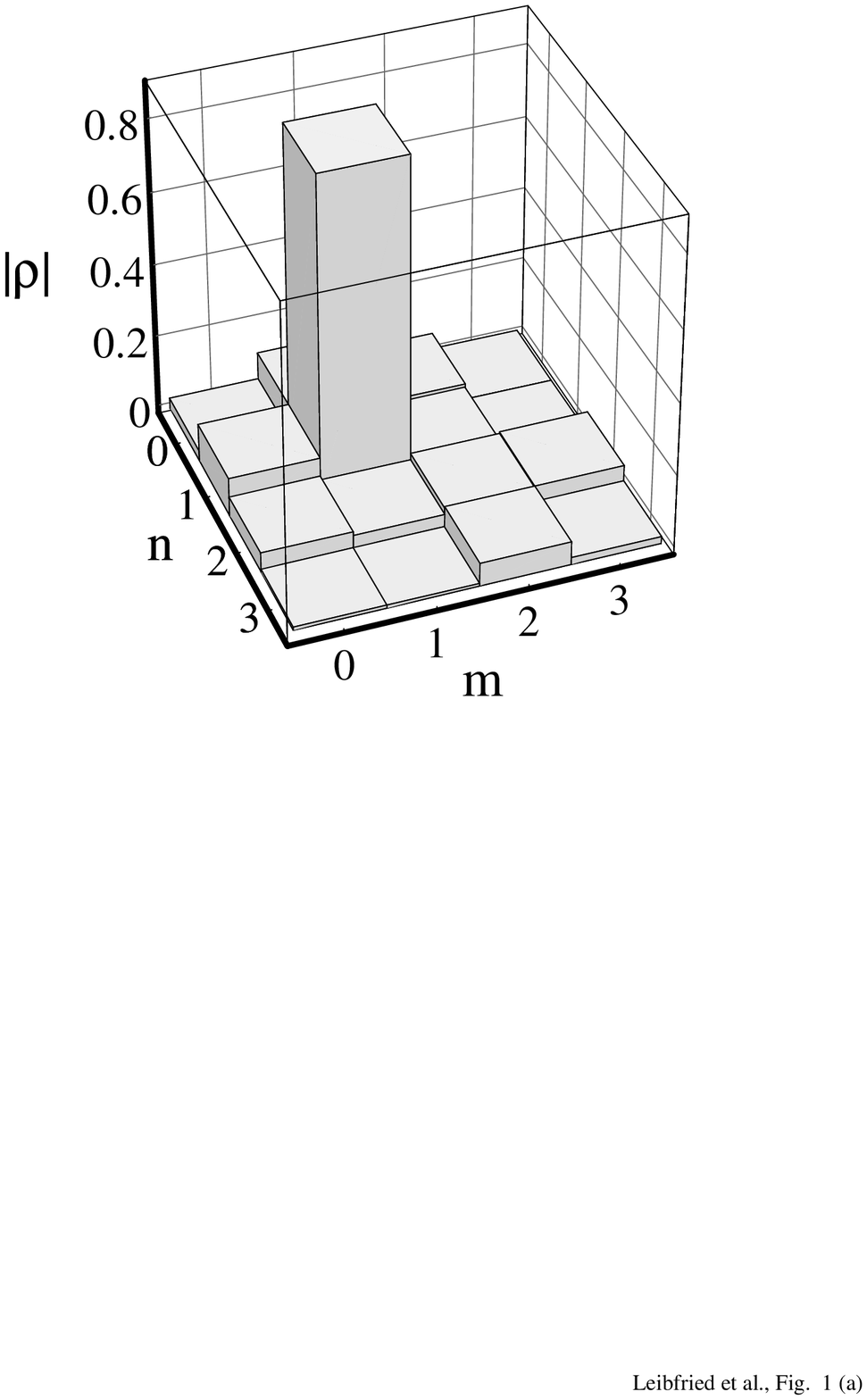}
}
\hspace{.5cm}
\subfigure[]{\includegraphics[width=11cm,clip]{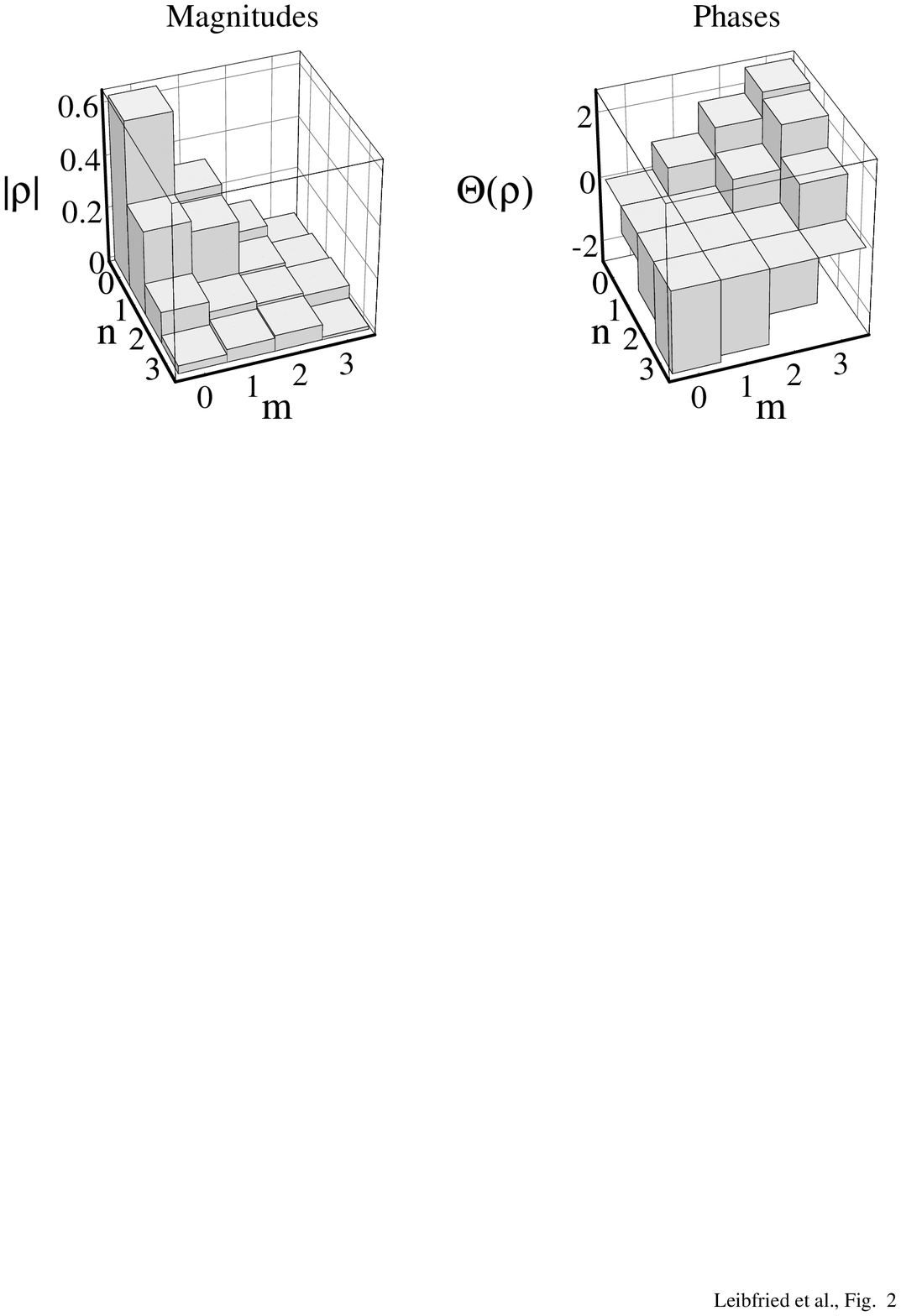}
}
\end{center}
\caption[n1rho]
{\label{fig:n1rho}
(a) Reconstructed number-state density matrix amplitudes $|\rho_{nm}|$ for
an approximate $n=1$ number state.
(b) Reconstructed amplitudes $|\rho_{nm}|$ and phases
$\Theta(\rho_{nm})$  of a coherent state.
}
\end{figure}

Figure \ref{fig:n1rho}(a) shows the reconstructed density matrix
amplitudes for an approximate $n=1$ state.
Figure \ref{fig:n1rho}(b) shows the reconstructed density matrix for a
coherent state having an amplitude $|\beta|\approx 0.67$.


\begin{figure}[tbhp]
\begin{center}
\includegraphics[width=11cm,clip]{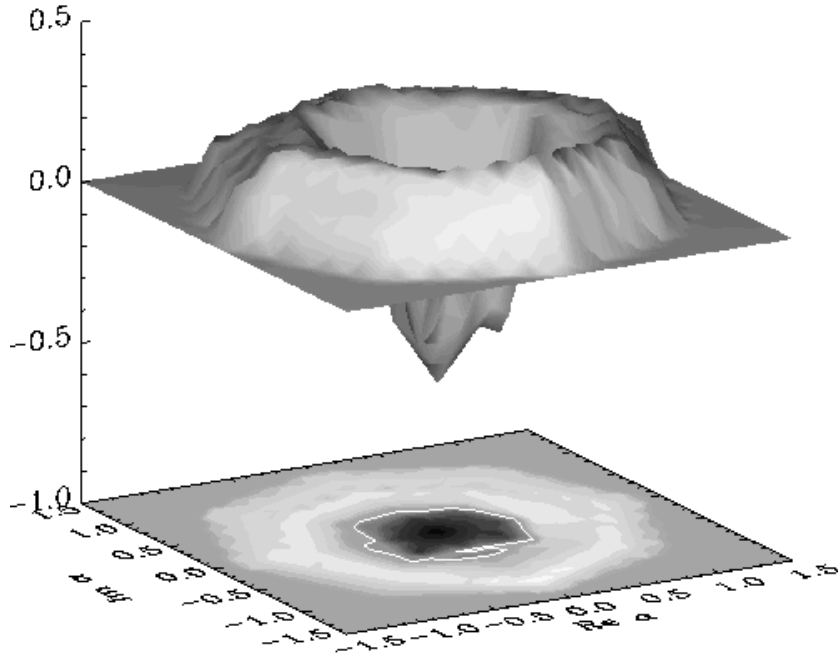}
\end{center}
\caption[n1wigner]
{\label{fig:n1wigner}
Surface and contour plots of the reconstructed
Wigner function $W(\alpha)$ of an approximate $n=1$ number state.
The negative values of $W(\alpha)$  around the origin highlight the
nonclassical nature of this state.
}
\end{figure}

\begin{figure}[tbhp]
\begin{center}
\includegraphics[width=11cm,clip]{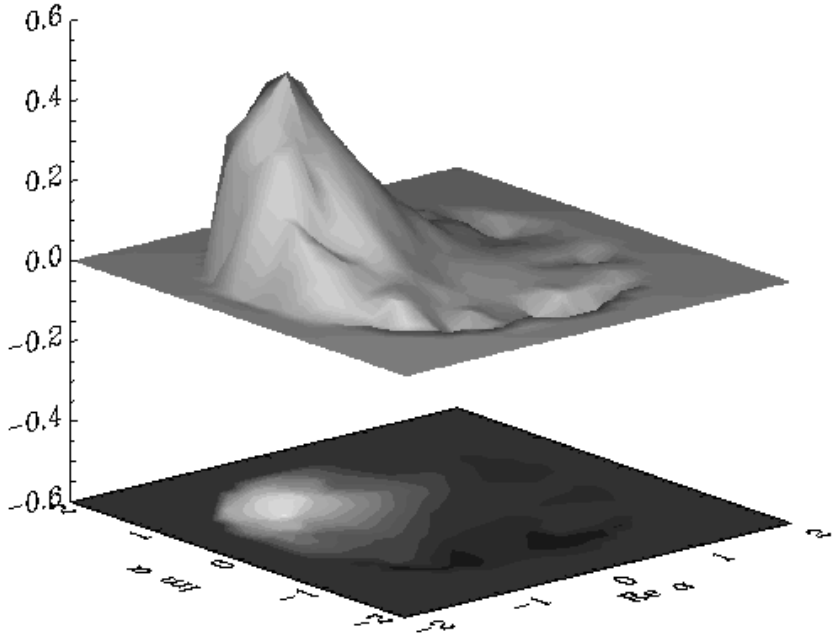}
\end{center}
\caption[cohwigfn]
{\label{fig:cohwigfn}
Surface and contour plots of the reconstructed
Wigner function $W(\alpha)$ of a coherent state.
The width is in good agreement with the expected minimum-uncertainty
value.
}
\end{figure}

The Wigner function for a given value of the complex parameter $\alpha$
can be determined from the sum
\cite{royer85,moya-cessa93,wallentowitz96,banaszek96}
\begin{equation}
W(\alpha)=\frac{2}{\pi}\sum_{n=0}^\infty (-1)^n Q_n(\alpha).
\end{equation}
Figure \ref{fig:n1wigner} shows the reconstructed Wigner function
for an approximate $n=1$ state.
The fact that it is negative in a region around the origin
highlights the fact that is a nonclassical state.
Figure \ref{fig:cohwigfn} shows the reconstructed Wigner function
for a coherent state with amplitude $|\beta|\approx 1.5$.
It is positive, which is not surprising, since the coherent state
is the quantum state that most closely approximates a classical
state.


\section{DECOHERENCE OF A SUPERPOSITION OF COHERENT STATES}

Decoherence, the decay of superposition states into mixtures,
is of interest because it leads to the familiar classical physics
of macroscopic objects (see, e.g. Ref.~\citenum{zurek91}).
Experiments are beginning to probe the regime in which decoherence
is observable, but not so fast that the dynamics are classical.
One example is the observation of the decoherence of a superposition
of mesoscopic (few-photon)
quantum states of an electromagnetic field having distinct phases
by Brune {\em et al.} \cite{brune96}
In this experiment, as in the Schr\"odinger cat experiment of
Ref.~\citenum{monroe96}, the mesoscopic system was entangled with
an internal two-level system of an atom.

A simpler system, which has been well-studied theoretically
by others \cite{walls85,paz93,garraway94,goetsch95}, is
a superposition of two coherent states of a harmonic oscillator,
without entanglement with a two-level system.
The system interacts in some way with the environment, leading
to decoherence of the superposition state.

Here we carry out a simple calculation of the decoherence of a superposition
of two coherent states of a harmonic oscillator.
At time $t=0$, the system is in an equal superposition of two coherent
states,
\begin{equation}
\label{eq:decoherence-0}
|\Psi(0)\rangle=\frac{1}{\sqrt{2}}\left(e^{i\theta_1(0)}|\alpha_1(0)\rangle
+e^{i\theta_2(0)}|\alpha_2(0)\rangle\right),
\end{equation}
where the notation is the same as in Appendix A, and
the interaction picture is used.
(It is assumed that the two components of the state vector are nearly
orthogonal.  Otherwise the normalization factor is different.)
The system is subjected to a random force, which is uniform over the spatial
extent of the system, for example, a uniform electric field acting on a
charged particle.
In Appendix A, we show that a single coherent state, subjected
to such a force, will remain in a coherent state.
It follows from the linearity of the Schr\"odinger equation that  the
state described by Eq.~(\ref{eq:decoherence-0}) will, at a later time,
still be in an equal superposition of coherent
states, although $\alpha_1$, $\alpha_2$, $\theta_1$, and $\theta_2$
will change.
Each time the system of Eq.~(\ref{eq:decoherence-0}) is prepared,
$\alpha_1$, $\alpha_2$, $\theta_1$, and $\theta_2$ will have
a different time dependence, due to the random force.
We can study the decoherence as a function of time by carrying out
an ensemble average.

Although both $\alpha_1$ and $\alpha_2$ change with time, their
difference does not, since they each change by the same amount.
This can be seen from Eq.~(\ref{eq:alpha}).
We define $\Delta\alpha\equiv \alpha_1-\alpha_2={\rm constant}$.

Decoherence is due to the changes in the phase difference
$[\theta_2(t)-\theta_1(t)]$.
We define the change in the phase difference to be
\begin{equation}
\Delta\theta(t)\equiv [\theta_2(t)-\theta_1(t)]-[\theta_2(0)-\theta_1(0)].
\end{equation}
Then, from Eq.~(\ref{eq:thetaeqn3}),
\begin{equation}
\Delta\theta(t)=\frac{1}{2}\left[\Delta\alpha
\int_0^tf(t^\prime)e^{-i\omega_xt^{\prime}}dt^\prime
+\Delta\alpha^*
\int_0^tf(t^\prime)e^{i\omega_xt^{\prime}}dt^\prime
\right].
\end{equation}
The double-integral term in Eq.~(\ref{eq:thetaeqn3})
does not contribute
to $\Delta\theta(t)$, since it does not depend on either $\alpha(0)$
or $\theta(0)$, so it makes the same contribution to
$\theta_2(t)$ and to $\theta_1(t)$.
The square of $\Delta\theta(t)$ is
\begin{eqnarray}
\left(\Delta\theta(t)\right)^2&=&
\frac{1}{4}\left[(\Delta\alpha)^2\int_0^t\int_0^t f(t^\prime)f(t^{\prime\prime})
e^{-i\omega_xt^\prime-i\omega_xt^{\prime\prime}}dt^\prime dt^{\prime\prime}
+2|\Delta\alpha|^2\int_0^t\int_0^t f(t^\prime)f(t^{\prime\prime})
e^{i\omega_xt^\prime-i\omega_xt^{\prime\prime}}dt^\prime dt^{\prime\prime}
\right.\nonumber\\
& &\left.
\mbox{}+(\Delta\alpha^*)^2\int_0^t\int_0^t f(t^\prime)f(t^{\prime\prime})
e^{i\omega_xt^\prime+i\omega_xt^{\prime\prime}}dt^\prime dt^{\prime\prime}
\right],
\end{eqnarray}
and its ensemble average is
\begin{eqnarray}
\left\langle\left(\Delta\theta(t)\right)^2\right\rangle&=&
\frac{1}{4}\left[(\Delta\alpha)^2\int_0^t\int_0^t
\langle f(t^\prime)f(t^{\prime\prime})\rangle
e^{-i\omega_xt^\prime-i\omega_xt^{\prime\prime}}dt^\prime dt^{\prime\prime}
+2|\Delta\alpha|^2\int_0^t\int_0^t
\langle f(t^\prime)f(t^{\prime\prime})\rangle
e^{i\omega_xt^\prime-i\omega_xt^{\prime\prime}}dt^\prime dt^{\prime\prime}
\right.\nonumber\\
& &\left.
\mbox{}+(\Delta\alpha^*)^2\int_0^t\int_0^t
\langle f(t^\prime)f(t^{\prime\prime})\rangle
e^{i\omega_xt^\prime+i\omega_xt^{\prime\prime}}dt^\prime dt^{\prime\prime}
\right],
\end{eqnarray}
where the angular brackets denote an ensemble average.
Thus, $\left\langle\left(\Delta\theta(t)\right)^2\right\rangle$
depends on the autocorrelation function
of the random function $f(t)$, which is
proportional to the force.
If we assume that $f$ is a stationary random variable, then its
autocorrelation function
\begin{equation}
R(\tau)=\langle f(t)f(t+\tau)\rangle
\end{equation}
exists and is independent of $t$.
According to the Wiener-Khinchine theorem, the power spectrum $w(\nu)$,
defined for $\nu\geq 0$, is
\begin{equation}
w(\nu)=4\int_0^\infty R(\tau)\cos(2\pi \nu\tau)d\tau.
\end{equation}
A white-noise spectrum corresponds to $R(\tau)\propto \delta(\tau)$,
where $\delta(\tau)$ is the Dirac delta function.
If we let $R(\tau)=C\delta(\tau)$, then
\begin{eqnarray}
\left\langle\left(\Delta\theta(t)\right)^2\right\rangle&=&
\frac{C}{4}\left[(\Delta\alpha)^2\int_0^t\int_0^t
\delta(t^{\prime\prime}-t^\prime)
e^{-i\omega_xt^\prime-i\omega_xt^{\prime\prime}}dt^\prime dt^{\prime\prime}
+2|\Delta\alpha|^2\int_0^t\int_0^t \delta(t^{\prime\prime}-t^\prime)
e^{i\omega_xt^\prime-i\omega_xt^{\prime\prime}}dt^\prime dt^{\prime\prime}
\right.\nonumber\\
& &\left.
\mbox{}+(\Delta\alpha^*)^2\int_0^t\int_0^t \delta(t^{\prime\prime}-t^\prime)
e^{i\omega_xt^\prime+i\omega_xt^{\prime\prime}}dt^\prime dt^{\prime\prime}
\right]\nonumber\\
&=&\frac{C}{4}\left[(\Delta\alpha)^2\int_0^t e^{-2i\omega_xt^\prime}dt^\prime
+2|\Delta\alpha|^2\int_0^t dt^\prime
+(\Delta\alpha^*)^2\int_0^t e^{2i\omega_xt^\prime}dt^\prime \right].
\label{eq:delta-theta1}
\end{eqnarray}
The first and third integrals on the right-hand-side of
Eq.~(\ref{eq:delta-theta1}) oscillate, but are bounded, while the second
one grows with time, so, for $\omega_x t \gg 1$,
\begin{equation}
\label{eq:delta-theta2}
\left\langle\left(\Delta\theta(t)\right)^2\right\rangle \approx
\frac{1}{2}C|\Delta\alpha|^2t.
\end{equation}
Hence, the decoherence time,
that is, the time required for the rms phase difference
$\left\langle\left(\Delta\theta(t)\right)^2\right\rangle^{1/2}$
to increase to about 1 radian, is on the order of
$2/\left(C|\Delta\alpha|^2\right)$.

Using the same approximations, we can calculate the
rate of change of the amplitude $\alpha_i(t)$ for a single coherent
state, where $i$ is 1 or 2.
From Eq.~(\ref{eq:alpha}) in Appendix A,
\begin{equation}
\alpha_i(t)-\alpha_i(0)=i\int_0^t f(t^\prime)e^{i\omega_xt^\prime}dt^\prime,
\end{equation}
and
\begin{equation}
|\alpha_i(t)-\alpha_i(0)|^2=\int_0^t \int_0^t f(t^\prime) f(t^{\prime\prime})
e^{i\omega_xt^\prime -i\omega_xt^{\prime\prime}} dt^\prime dt^{\prime\prime}.
\end{equation}
If $f(t)$ has a white-noise spectrum, as considered previously,
the ensemble average is
\begin{equation}
\left\langle|\alpha_i(t)-\alpha_i(0)|^2\right\rangle
=C\int_0^t \int_0^t \delta(t^{\prime\prime}-t^\prime)
e^{i\omega_xt^\prime -i\omega_xt^{\prime\prime}} dt^\prime dt^{\prime\prime}
=C\int_0^t dt^\prime =Ct.
\end{equation}
In the time, $t\approx 2/\left(C|\Delta\alpha|^2\right)$,
required for decoherence, the rms change in $\alpha_i$ is
$\sqrt{2}/|\Delta\alpha|$.
Consider the case $\alpha_1(0)=-\alpha_2(0)$.
The energy of a single coherent state is proportional to the square
of its amplitude ($\propto |\alpha_i|^2$).
The fractional change in the energy is
\begin{equation}
\left|\frac{\Delta|\alpha_i|^2}{|\alpha_i|^2}\right|
<\frac{2|\alpha_i(t)-\alpha_i(0)|}{\alpha_i(0)}
\approx \frac{2\sqrt{2}}{|\Delta\alpha||\alpha_i(0)|}
=\frac{\sqrt{2}}{|\alpha_i(0)|^2},
\end{equation}
which becomes small for $|\alpha_i(0)|^2\gg 1$.
Thus, we see that coherent superpositions of macroscopic
($|\alpha_i(0)|^2\gg 1$) states decohere much more quickly than
they change in energy.

We can make a more quantitative statement about the decoherence
rate by considering the decay of the off-diagonal
density matrix elements.
For an initial pure state $2^{-1/2}(|\alpha_1\rangle+|\alpha_2\rangle)$,
the density matrix is
\begin{equation}
\rho=\frac{1}{2}\left(|\alpha_1\rangle\langle\alpha_1|
+|\alpha_1\rangle\langle\alpha_2|
+|\alpha_2\rangle\langle\alpha_1|
+|\alpha_2\rangle\langle\alpha_2|\right).
\end{equation}
If $\Delta\alpha\gg 1$, so that the phase changes are more important
than the changes in $\alpha_1$ and $\alpha_2$,
then the off-diagonal matrix element
$|\alpha_2\rangle\langle\alpha_1|$ evolves to
$e^{i\Delta\theta(t)}|\alpha_2\rangle\langle\alpha_1|$.
Let us assume that the random variable $\Delta\theta$ at time $t$
has a Gaussian distribution $P(\Delta\theta)$.
Given the variance of $\Delta\theta$ from Eq.~(\ref{eq:delta-theta2}),
$P(\Delta\theta)$ must have the form
\begin{equation}
P(\Delta\theta)=\frac{1}{\sqrt{\pi C|\Delta\alpha|^2t}}
\exp\left(-\frac{\Delta\theta^2}{C|\Delta\alpha|^2t} \right).
\end{equation}
We evaluate the ensemble average of $e^{i\Delta\theta(t)}$,
\begin{equation}
\left\langle e^{i\Delta\theta(t)} \right\rangle=
\int_{-\infty}^{\infty} P(\Delta\theta) e^{i\Delta\theta(t)} d(\Delta\theta)
=\exp\left(-\frac{C|\Delta\alpha|^2t}{4}\right).
\end{equation}
This yields the basic result, previously derived by others
\cite{walls85,paz93,garraway94,goetsch95},
that the off-diagonal matrix element decays exponentially
in time, with a time constant that is inversely proportional to
$|\Delta\alpha|^2$.

An extension of the experiment of Ref.~\citenum{monroe96}, with
a random electric field deliberately applied, might be used to
verify these calculations \cite{monroe96}.
A means of simulating thermal, zero-temperature, squeezed,
and other reservoirs by various combinations of optical fields
has been discussed by Poyatos {\em et al.} \cite{poyatos96}


\section*{ACKNOWLEDGMENTS}

This work was supported by the National Security Agency, the Army
Research Office, and the Office of Naval Research.
D.\ L.\ acknowledges a Deutsche Forschungsgemeinschaft research grant.
D.\ M.\ M.\ was supported by a National Research Council postdoctoral
fellowship.

\appendix

\section{THE FORCED HARMONIC OSCILLATOR}
\label{sec:forced}

If a coherent state is subjected to a spatially uniform force,
it remains a coherent state, though its amplitude changes.
This exact result, which is independent of the strength or time-dependence
of the force, seems to have been discovered by Husimi \cite{husimi53}
and independently by Kerner \cite{kerner58}.
For other references on the forced quantum harmonic oscillator, see
Ref.~\citenum{nogami91}.
The qualitative result stated above is enough to establish that
applying a force to the $n=0$ state will generate a finite-amplitude
coherent state, which was required in
Sec.~\ref{subsec:coherent}.~~
However, in order to study the decoherence of a superposition of two
coherent states, it is useful to have an exact expression for the
time-dependent state vector, given that the state vector is
an arbitrary coherent state at $t=0$.
Since the published results of which we are aware do not give this
particular result, we give here an elementary derivation.

We consider a Hamiltonian, $H=H_0+V(t)$,
where $H_0=\hbar\omega_x a_x^\dagger a_x$
is the Hamiltonian of a one-dimensional harmonic oscillator,
and $V(t)=-xF(t)$ is a time-dependent potential.
Here, $F(t)$ is a real $c$-number function of time
and corresponds to a spatially uniform force,
and $x=x_0(a_x+a_x^\dagger)$, where $x_0\equiv (\hbar/2m\omega_x)^{1/2}$.
For example, if the particle has charge $q$, and a uniform electric
field $\hat{x}E_x(t)$ is applied, then $F(t)=qE_x(t)$.

It is convenient to switch to the interaction picture, where
an interaction-picture state vector $|\Psi(t)\rangle^I$ is related
to the corresponding Schr\"odinger-picture state vector
$|\Psi(t)\rangle^S$ by
\begin{equation}
|\Psi(t)\rangle^I =e^{iH_0 t/\hbar}|\Psi(t)\rangle^S,
\end{equation}
and an interaction-picture operator ${\cal O}^I(t)$ is related to the
corresponding Schr\"odinger-picture operator ${\cal O}^S(t)$ by
\begin{equation}
{\cal O}^I(t) =e^{iH_0 t/\hbar}{\cal O}^S(t)e^{-iH_0 t/\hbar}.
\end{equation}
The equation of motion for an interaction-picture operator is
\begin{equation}
\frac{d}{dt}{\cal O}^I(t)=-\frac{i}{\hbar}\left[{\cal O}^I(t),H_0 \right],
\end{equation}
so
\begin{eqnarray}
\frac{d}{dt}a_x^I(t)&=&-\frac{i}{\hbar}\left[a_x^I(t),H_0\right]=
-i\omega_x a^I(t),\\
a_x^I(t)&=&a_x^I(0)e^{-i\omega_x t},\\
{a_x^\dagger}^I(t)&=&{a_x^\dagger}^I(0)e^{i\omega_x t}.
\end{eqnarray}
For simplicity, we let
$a\equiv a_x^I(0)=a_x^S$ and
$a^\dagger\equiv {a_x^\dagger}^I(0)={a_x^\dagger}^S$.
The Schr\"odinger equation in the interaction picture is
\begin{equation}
\label{eq:schrod}
\frac{\partial}{\partial t}|\Psi(t)\rangle^I=
-\frac{i}{\hbar}V^I(t)|\Psi(t)\rangle^I
=if(t)\left(ae^{-i\omega_xt}+a^\dagger e^{i\omega_x t} \right)
|\Psi(t)\rangle^I,
\end{equation}
where $f(t)\equiv x_0 F(t)/\hbar$.
Note that if $f(t)=0$, $|\Psi(t)\rangle^I$ is constant.

We impose the initial condition that $|\Psi(0)\rangle^I$ is a coherent
state.
We make the assumption that $|\Psi(t)\rangle^I$ is also a coherent
state for $t>0$,
\begin{equation}
\label{eq:ansatz}
|\Psi(t)\rangle^I=e^{i\theta(t)}|\alpha(t)\rangle
\equiv e^{i\theta(t)}e^{-\frac{1}{2}|\alpha(t)|^2}
\sum_{n=0}^\infty\frac{\alpha^n(t)}{\sqrt{n!}}|n\rangle,
\end{equation}
where the $|n\rangle$ are the
eigenstates of $a_x^\dagger a_x$ with eigenvalue $n$ at $t=0$,
and $\theta(t)$ is an arbitrary real function, so that
$^I\langle\Psi(t)|\Psi(t)\rangle^I=1$.
(This assumption
must be verified later by substitution of the resulting solution
back into Eq.~(\ref{eq:schrod}).)
The problem reduces to finding the complex function $\alpha(t)$
and the real function $\theta(t)$.

The right-hand-side of Eq.~(\ref{eq:schrod}) is
\begin{eqnarray}
-\frac{i}{\hbar}V^I(t)|\Psi(t)\rangle^I&=&if(t)
\left(ae^{-i\omega_xt}+a^\dagger e^{i\omega_x t} \right)
e^{i\theta(t)}e^{-\frac{1}{2}|\alpha(t)|^2}
\sum_{n=0}^\infty\frac{\alpha^n(t)}{\sqrt{n!}}|n\rangle\nonumber\\
&=&if(t)e^{i\theta(t)}e^{-\frac{1}{2}|\alpha(t)|^2}
\left( e^{-i\omega_xt}\sum_{n=1}^\infty\frac{\alpha^n(t)\sqrt{n}}{\sqrt{n!}}
|n-1\rangle
+ e^{i\omega_xt}\sum_{n=0}^\infty\frac{\alpha^n(t)\sqrt{n+1}}{\sqrt{n!}}
|n+1\rangle\right)\label{eq:rhs}\\
&=&if(t)e^{i\theta(t)}e^{-\frac{1}{2}|\alpha(t)|^2}
\left( e^{-i\omega_xt}\sum_{n=0}^\infty\frac{\alpha^{n+1}(t)}{\sqrt{n!}}
|n\rangle
+ e^{i\omega_xt}\sum_{n=1}^\infty\frac{\alpha^{n-1}(t)\sqrt{n}}{\sqrt{(n-1)!}}
|n\rangle\right).\nonumber
\end{eqnarray}
The left-hand-side of Eq.~(\ref{eq:schrod}) is
\begin{eqnarray}
\frac{\partial}{\partial t}|\Psi(t)\rangle^I&=&
i\frac{d\theta}{dt}e^{i\theta(t)}e^{-\frac{1}{2}|\alpha(t)|^2}
\sum_{n=0}^\infty\frac{\alpha^{n}(t)}{\sqrt{n!}}|n\rangle
+e^{i\theta(t)}\left(-\frac{1}{2}\frac{d\alpha}{dt}\alpha^*(t)
-\frac{1}{2}\alpha(t)\frac{d\alpha^*}{dt}
\right)e^{-\frac{1}{2}|\alpha(t)|^2}
\sum_{n=0}^\infty\frac{\alpha^{n}(t)}{\sqrt{n!}}|n\rangle\nonumber\\
& &\mbox{}+ie^{i\theta(t)}e^{-\frac{1}{2}|\alpha(t)|^2}
\sum_{n=0}^\infty\frac{n\alpha^{n-1}(t)}{\sqrt{n!}}\frac{d\alpha}{dt}
|n\rangle.\label{eq:lhs}
\end{eqnarray}
We equate the coefficients of $|n\rangle$ in Eqs.~(\ref{eq:rhs}) and
(\ref{eq:lhs}) and divide by a common factor to obtain
\begin{equation}
if(t)\left( e^{-i\omega_xt}\alpha^2(t)+e^{i\omega_xt}n\right)
=i\frac{d\theta}{dt}\alpha(t)
+\left(-\frac{1}{2}\frac{d\alpha}{dt}\alpha^*(t)
-\frac{1}{2}\alpha(t)\frac{d\alpha^*}{dt}
\right)\alpha(t) +n\frac{d\alpha}{dt},
\end{equation}
which can be rearranged as
\begin{equation}
\label{eq:n-coeff}
n\left[ \frac{d\alpha}{dt} -if(t)e^{i\omega_xt}\right]
+i\alpha(t)\frac{d\theta}{dt}
-\frac{1}{2}\alpha(t)\alpha^*(t)\frac{d\alpha}{dt}
-\frac{1}{2}\alpha^2(t)\frac{d\alpha^*}{dt}
-i\alpha^2(t)f(t)e^{-\omega_xt}=0
\end{equation}
In order for Eq.~(\ref{eq:n-coeff}) to be true for all $n$,
the expression in the square brackets, which multiplies $n$,
must equal zero.
Thus, we have the first-order differential equation for $\alpha(t)$,
\begin{equation}
\label{eq:alphaeqn}
\frac{d\alpha}{dt}-if(t)e^{i\omega_xt}=0,
\end{equation}
which has the solution
\begin{equation}
\label{eq:alpha}
\alpha(t)=\alpha(0)+i\int_0^t f(t^\prime)e^{i\omega_xt^\prime}dt^\prime.
\end{equation}

If the expression in brackets is set to zero, Eq.~(\ref{eq:n-coeff})
becomes, after division of both sides by $i\alpha(t)$,
\begin{equation}
\label{eq:thetaeqn}
\frac{d\theta}{dt}+\frac{i}{2}\alpha^*(t)\frac{d\alpha}{dt}
+\frac{i}{2}\alpha(t)\frac{d\alpha^*}{dt} - \alpha(t)f(t)e^{-i\omega_xt}
=0.
\end{equation}
The imaginary part of Eq.~(\ref{eq:thetaeqn}) is satisfied as long as
Eq.~(\ref{eq:alphaeqn}) is, so it provides no new information.
The real part of Eq.~(\ref{eq:thetaeqn}) is a first-order differential
equation for $\theta(t)$:
\begin{eqnarray}
\frac{d\theta}{dt}&=&f(t)\left[{\rm Re}\,\alpha(t)\cos\omega_xt
+{\rm Im}\,\alpha(t) \sin\omega_xt \right]\nonumber\\
&=&f(t)\left[{\rm Re}\,\alpha(0)\cos\omega_xt+{\rm Im}\,\alpha(0)\sin\omega_xt
-\cos\omega_xt\int_0^t f(t^\prime)\sin\omega_xt^\prime dt^\prime
+\sin\omega_xt\int_0^tf(t^\prime)\cos\omega_xt^\prime dt^\prime
\right]\label{eq:theta2eqn}\nonumber\\
&=&f(t)\left[{\rm Re}\,\alpha(0)\cos\omega_xt+{\rm Im}\,\alpha(0)\sin\omega_xt
+ \int_0^t f(t^\prime)\sin\left[\omega_x(t-t^\prime)\right] dt^\prime\right],
\label{eq:thetaeqn2}\end{eqnarray}
where Re and Im stand for the real and imaginary parts.
Integrating Eq.~(\ref{eq:thetaeqn2}), we obtain
\begin{equation}
\theta(t)=\theta(0)+{\rm Re}\,\alpha(0)\int_0^tf(t^\prime)\cos\omega_x
t^\prime dt^\prime + {\rm Im}\,\alpha(0)\int_0^t f(t^\prime)\sin\omega_x
t^\prime dt^\prime
+\int_0^t dt^\prime f(t^\prime)\int_0^{t^\prime} dt^{\prime\prime}
f(t^{\prime\prime})\sin\left[\omega_x(t^\prime-t^{\prime\prime}) \right].
\label{eq:thetaeqn3}\end{equation}
Substitution back into the Schr\"odinger equation [Eq.~(\ref{eq:schrod})]
verifies the solution and justifies the original
assumption [Eq.~(\ref{eq:ansatz})].


\bibliography{spie03}   

\bibliographystyle{spiebib}   

\end{document}